\documentclass[preprint,showpacs,preprintnumbers,amsmath,amssymb, superscriptaddress,longbibliography,nofootinbib,aps,prd]{revtex4-1}
\usepackage{graphicx} % Required for inserting images
\usepackage{xcolor}
\usepackage{amsmath}
 \usepackage{hyperref}
\usepackage[top = 2cm, bottom = 2cm, right = 2cm, left =2cm]{geometry}

\begin{document}

\title{Field Sources for Generalized Ellis-Bronnikov Wormhole}

\author{T. M. Crispim \footnote{Author to whom any correspondence should be addressed.}\footnote{E-mail: tiago.crispim@fisica.ufc.br}}\affiliation{Departamento de F\'isica, Universidade Federal do Cear\'a, Caixa Postal 6030, Campus do Pici, 60455-760 Fortaleza, Cear\'a, Brazil}

\author{G. Alencar \footnote{E-mail: geova@fisica.ufc.br}}\affiliation{Departamento de F\'isica, Universidade Federal do Cear\'a, Caixa Postal 6030, Campus do Pici, 60455-760 Fortaleza, Cear\'a, Brazil}

\author{C. R. Muniz\footnote{E-mail: celio.muniz@uece.br}}\affiliation{Universidade Estadual do Cear\'a (UECE), Faculdade de Educa\c c\~ao, Ci\^encias e Letras de Iguatu, 63500-000, Iguatu, CE, Brazil.}

\date{\today}

\begin{abstract}
The so-called generalized Ellis–Bronnikov wormhole is a modification of the standard Ellis–Bronnikov solution, in which a parameter 
$m>2$ is introduced-recovering the original Ellis-Bronnikov geometry when 
$m=2$. In this work, we investigate the properties of this spacetime by analyzing its embedding diagrams and how they are affected by variations in the parameter 
$m$. Furthermore, we study the accretion of dust onto this geometry, showing that, unlike in black hole scenarios, the radial infall velocity of the dust decreases as it approaches the wormhole throat, with this deceleration becoming increasingly abrupt for larger values of $m$. Our results also demonstrate that the mass of the wormhole generally decreases due to the accretion process, a finding that aligns with recent works in the literature for Ellis-Bronnikov-type geometries. This mass loss, coupled with the characteristic accumulation of matter near the throat, highlights the unique dynamical response of traversable wormholes to baryonic influx. As a main result, we demonstrate that this geometry arises as an exact solution of General Relativity when considering the combined presence of a phantom scalar field and a magnetic or electric source.
\end{abstract}

\maketitle

\section{Introduction}

In the context of General Relativity (GR), wormholes are exact solutions of Einstein's equation that describe geometries characterized by the existence of a tunnel capable of connecting two different regions of a spacetime or even two different spacetimes through its throat \cite{Morris1988WormholesIS}. Although, like black holes, they are spherically symmetric solutions, wormholes are distinguished by being free of event horizons. Moreover, it is known from the literature that the shadows of black holes and horizonless objects, such as naked singularities \cite{Bambi:2008jg,Ortiz_2015,Shaikh_2018} and wormholes \cite{Bambi:2013nla,Azreg_A_nou_2015,Ishkaeva:2023xny,Alloqulov:2024olb,Guerrero:2021ues}, can be similar in some cases \cite{Shaikh:2018kfv}, which, along with the release of the first black hole shadow image by the Event Horizon Telescope (EHT) \cite{EventHorizonTelescope:2019dse,EventHorizonTelescope:2022wkp}, has revitalized interest in studying such objects. In this context, recent works have investigated detection methods for wormholes in both general relativity and alternative theories of gravity, emphasizing their potential astrophysical significance \cite{DeFalco:2020afv,DeFalco:2023twb,Jana:2024pdh}.

It is well-established in the literature that exotic matter, i.e., matter that violates the energy conditions, is required to support such a traversable geometry. In this context, the simplest solution for a traversable Lorentzian wormhole is the Ellis-Bronnikov (EB) wormhole \cite{Ellis:1973yv,Bronnikov:1973fh}, whose metric is given by
\begin{equation}\label{eb}
    ds^2 = - dt^2 + dl^2 + (l^2 + b_0^2)d\Omega^2,
\end{equation}
where $l$ is known as the proper radial distance, and $b_0$ is the radius of the wormhole's throat. It is generated by an action where gravity is minimally coupled to a free phantom scalar field \cite{Bambi:2008jg}
\begin{equation}
    S =  \int d^4x \left(R - 2\epsilon g^{\mu\nu}\partial_\mu\phi\partial_\nu\phi \right),
\end{equation}
where $\epsilon = -1$ (phantom) and $\phi = \phi(l)$ is given by
\begin{equation}
    \phi(l) = \phi_0 + \arctan\left(\frac{l}{b_0}\right).
\end{equation}

The EB metric is the simplest one describing a traversable wormhole that satisfies the conditions presented in \cite{Morris1988WormholesIS}. However, in \cite{Kar:1995jz}, a generalization of the EB metric is introduced by incorporating a free parameter $m \geq 2$, such that
\begin{equation}\label{eb1}
    ds^2 = - dt^2 + dl^2 + (l^m + b_0^m)^{2/m}d\Omega^2,
\end{equation}
where $m = 2$  recovers the EB wormhole. The following section will discuss some characteristics of this wormhole family.

On the other hand, the metrics of so-called black bounces \cite{Simpson:2018tsi,Franzin:2021vnj}, i.e., spacetimes that transition from regular black holes to wormholes, have been extensively studied recently, with the Simpson-Visser procedure also being applied in other contexts \cite{Crispim_2024,Lima:2023jtl,Lima:2022pvc, Furtado:2022tnb, Pereira:2025fvg, Crispim:2024nou}. An interesting aspect of these geometries is that the field sources generating them are generally composed of Nonlinear Electrodynamics (NED) combined with phantom scalar fields \cite{Bronnikov:2024izh}, which has driven the search for field sources of black bounce geometries \cite{Bronnikov:2021uta,Alencar:2024yvh,Lima:2023arg,Bronnikov:2023aya,Rodrigues:2023vtm, Alencar:2025jvl} and regular black holes \cite{Bardeen:1968,Rodrigues:2018bdc} to become an actively explored topic recently. The NED theories modify the standard linear behavior of electromagnetic fields, especially in strong-field regimes, and have been explored in different contexts. In the context of wormholes, recent works have presented methods for constructing traversable wormholes supported by NED \cite{Canate:2022dzb}, as well as techniques for producing traversable wormholes with electric and magnetic charges without requiring exotic matter \cite{Canate:2024lks}.

Thus, this work investigates whether scalar fields combined with NED can serve as field sources for Generalized Ellis-Bronnikov (GEB) spacetimes. In the context of NED theories, we employ both magnetic and electric sources, having found analytical expressions for the scalar field, the associated potential, and the Lagrangian of the NED field. In the case of the magnetic solution, we were able to invert the equations to explicitly write the Lagrangian as a function of the invariant $\mathcal{F} = F^{\mu\nu}F_{\mu\nu}$.

This work is structured as follows: In the next section, a brief review of the main characteristics of the GEB metric will be provided, where we analyze the embedding diagrams of the geometry and how dust accretion is affected by the geometry; the following section presents our solution, both for the magnetic and electric cases; finally, in the last section, we will offer our discussion and final remarks.

\section{Generalized Ellis–Bronnikov (GEB) Space-time}

\subsection{Embedding diagram}
As previously mentioned, the EB wormhole metric is a solution to Einstein's field equation, where a massless scalar field with negative energy acts as the source of curvature. The metric is typically expressed as
\begin{equation}\label{metrica}
    ds^2 = -dt^2 + \frac{dr^2}{\left(1 - \frac{b(r)}{r}\right)} + r^2d\Omega^2,
\end{equation}
witch corresponds to the wormhole
 metric introduced by Morrison and Thorne \cite{Morris1988WormholesIS}, with a trivial redshift function.
Here, $b(r)$ is the shape function that governs the behavior of the wormhole.

The non-zero components of the Einstein tensor for the metric \eqref{metrica} are
\begin{equation}\label{Et}
    G^t_t = -\frac{b'(r)}{r^2},
\end{equation}
\begin{equation}\label{Er}
    G^r_r = - \frac{b(r)}{r^3},
\end{equation}
\begin{equation}\label{Ep}
    G^\theta_\theta = G^\varphi_\varphi  = \frac{b(r) - rb'(r)}{2r^3}.
\end{equation}

For the EB wormhole, the shape function takes on a particular form:
\begin{equation}
    b(r) = \frac{b_0^2}{r}
\end{equation}
where $b_0$ is the radius of the wormhole's throat. The geometry of the solution describes a ``tunnel" that connects two distinct regions of spacetime through the throat. This solution is one of the simplest types of wormhole solutions.

Through a change of coordinates given by
\begin{equation}
    dl^2 = \frac{dr^2}{\left(1 - \frac{b_0^2}{r^2}\right)}
\end{equation}
it is possible to express the metric as \eqref{eb}.

We can also express the GEB metric in terms of \( r \), where the metric takes the form given in \eqref{metrica}, where now
\begin{equation}\label{b}
    b(r) = r - r^{3 -2m}(r^m - b_0^m)^{2 - \frac{2}{m}}.
\end{equation}

Setting $t = \text{constant}$ and considering the spherical symmetry of the wormhole geometry, we can, without loss of generality, choose $\theta = \pi/2$ in Eq.~\eqref{eb1}. This yields the 2D spatial geometry described by the line element
\begin{equation}
    ds^2_{2D} = dl^2 + (l^m + b_0^m)^{2/m}d\varphi^2.
\end{equation}

We can visualize this curved 2D geometry by embedding it into a 3D Euclidean space with the line element
\begin{equation}
    d\sigma^2_{3D} = d\rho^2 + \rho^2d\varphi^2 + dz^2 = \left[\left(\frac{d\rho(l)}{dl}\right)^2 + \left(\frac{dz(l)}{dl}\right)^2\right]dl^2 + \rho^2d\varphi^2,
\end{equation}
where we identify the radial profile and the embedding coordinate $z(l)$ as
\begin{equation}
    \rho(l) = (b_0^m + l^m)^{1/m},
\end{equation}
\begin{equation}\label{zeq}
    \frac{dz}{dl} = \sqrt{1 - l^{-2 + 2m}(b_0^m + l^m)^{-2 + 2/m}}.
\end{equation}

\begin{figure}[ht]
    \centering
    \includegraphics[width=1\linewidth]{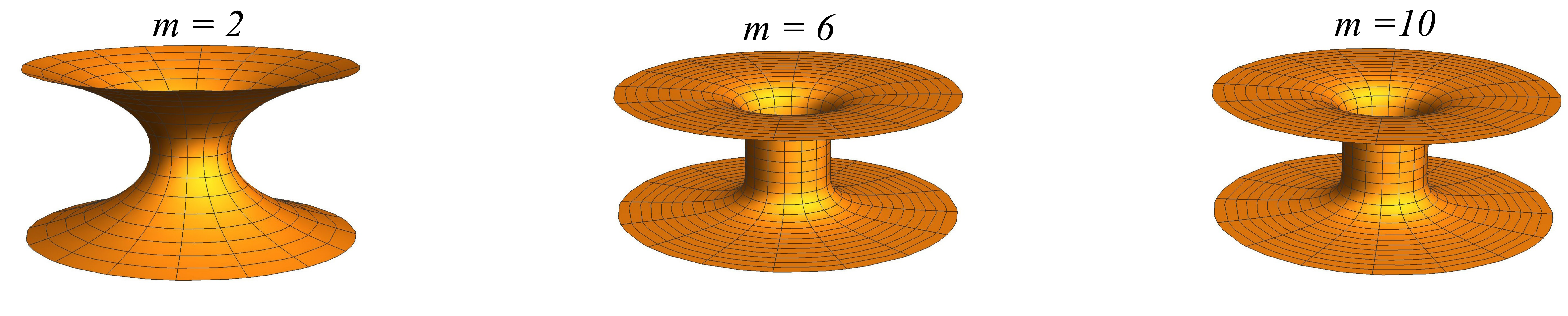}
    \caption{Embedding diagram for GEB wormhole for $b_0 = 1$ and different values of $m$.}
    \label{embebed}
\end{figure}

Numerically integrating Eq.~\eqref{zeq} with $b_0 = 1$ for different values of $m$ allows us to construct the embedding diagrams for the GEB wormhole, as shown in Fig.~\ref{embebed}. For the standard case ($m = 2$), the geometry exhibits the classical smooth shape of a catenoid. However, as $m$ increases ($m > 2$), the radial function $\rho(l)$ becomes exceedingly flat near $l \approx 0$. Consequently, the geometry of the wormhole's throat loses its simple catenoid shape and increasingly tends toward an elongated cylindrical tube. 

This geometric behavior is intimately related to the \textit{flaring-out condition}, a fundamental requirement for traversable wormholes \cite{Morris1988WormholesIS}. The flaring-out condition dictates that the inverse of the radial coordinate must be a minimum at the throat, which in terms of the embedding implies $d^2\rho/dz^2 > 0$. Physically, this geometry is sustained by the presence of exotic matter (in our case, the phantom scalar coupled with the NED field), which exerts an effective repulsive gravitational interaction to keep the throat open. For larger values of $m$, the required flaring-out is localized further away from the throat center, leaving a central cylindrical region with near-zero spatial curvature.

Interestingly, this geometric visualization provides a direct and simple explanation for the qualitative behavior of the infall velocity of the accreting dust discussed in the following subsection. Unlike a black hole scenario, where gravity is universally attractive and accelerates infalling matter towards the singularity, the effective repulsive geometry of the wormhole's throat acts as a geometric bottleneck. As the dust approaches the throat, it is decelerated by this repulsive effect. For higher values of $m$, the transition from the asymptotic flat region to the cylindrical throat becomes geometrically sharper. This abrupt spatial transition forces the infalling fluid to lose its radial velocity much more suddenly, directly reflecting the geometrical ``wall" of the elongated cylindrical neck seen in the embedding diagrams.

Energy conditions continue to be violated for this class of wormholes \cite{Kar:1995jz}. However, the geometry flattens more rapidly as $m$ increases, leading to resonances in the transmission of massless waves. This effect can be used to obtain information about the size of the wormhole's throat \cite{Kar:1995jz}. Furthermore, studies suggest that this family of wormholes can serve as a template for exploring the gravitational wave physics of exotic compact objects \cite{DuttaRoy:2019hij}. They may also act as potential black hole mimickers \cite{Roy:2021jjg}, provided stability under all types of perturbations is confirmed, a possibility supported by initial ringdown comparisons. Finally, this family of solutions has been extensively studied beyond GR, including braneworld models \cite{Sharma:2022tiv,Sharma:2022dbx,Sharma:2021kqb}, modified gravity \cite{Godani:2023jhq,Muniz:2022aal,Nilton:2022hrp}, and even condensed matter systems \cite{deSouza:2022ioq}.

\subsection{Accretion on GEB wormhole}

The accretion process of a fluid onto a black hole is already well documented in the literature\cite{Mach:2013gia,Yang:2015sfa,Bahamonde:2015uwa,Ahmed:2016cuy,Azreg-Ainou:2018wjx,Neves:2019ywx,Panotopoulos:2021ezt}. However, the study of accretion in wormhole geometries remains a relatively unexplored topic \cite{Rueda:2023val,Combi:2024ehi}, making the analysis of dust accretion (a pressureless fluid) particularly interesting to investigate in the wormhole geometry considered here.

The energy-momentum tensor for  dust is given by
\begin{equation}
    T_{\mu\nu} = \rho u_\mu u_\nu,
\end{equation}
where $\rho$ is the energy density and $u$ is the four-velocity of the fluid, defined as
\begin{equation}
    u^\mu \equiv \frac{d x^\mu}{d\tau} = (u^t, u^r,0,0)
\end{equation}
with proper time $\tau$, where from now on we will denote the radial velocity as $u^r \equiv u$. Here we will assume radial accretion, so that we are assuming $u^\theta = u^\varphi = 0$. From the normalization condition $u^\mu u_\mu = -1$, we can relate $u^t$ to $u^r$ as
\begin{equation}
    u^t = \sqrt{\frac{u^2}{\left(1 - \frac{b_0^m}{r^m}\right)^{2 - 2/m}} + 1}.
\end{equation}

On the other hand, the conservation equation of the energy-momentum tensor, $\nabla_\mu T^{\mu\nu} = 0$, implies
\begin{equation}
  \label{eq1acrecion}  \frac{\rho u r^2}{\left(1 - \frac{b_0^m}{r^m}\right)^{2 - 2/m}}\sqrt{u^2 + \left(1 - \frac{b_0^m}{r^m}\right)^{2 -2/m}}= C_1,
\end{equation}
where $C_1$ is an integration constant. Next, we will consider the equation for the conservation of mass flux, given by $\nabla_\mu J^\mu=0$, where $J^\mu = u_\nu T^{\mu\nu}$, which leads us to the expression
\begin{equation}
  \label{eq2acrecion}  \frac{\rho u r^2}{\sqrt{\left(1 - \frac{b_0^m}{r^m}\right)^{2 - 2/m}}} = C_2,
\end{equation}
where $C_2$ is another integration constant. By dividing the equations \eqref{eq1acrecion} and \eqref{eq2acrecion}, we obtain
\begin{equation}
    \sqrt{\frac{u^2}{\left(1 - \frac{b_0^m}{r^m}\right)^{2 -2/m}} + 1}=\frac{C_1}{C_2}\equiv C_3.
\end{equation}

\begin{figure}
    \centering
    \includegraphics[width=0.5\linewidth]{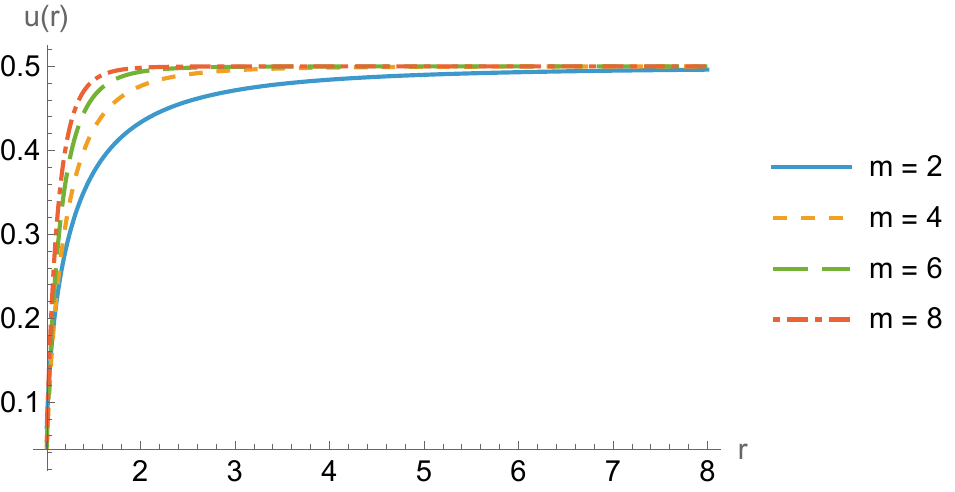}
    \caption{Radial velocity magnitude for different values of $m$ and fixing $b_0=1$.}
    \label{radial}
\end{figure}

With this, we can finally obtain expressions for $u$ and $\rho$
\begin{eqnarray}
     u &=& \pm \sqrt{\left(1 -\frac{b_0^m}{r^m}\right)^{2 -2/m}({C^{(m)}_3}^2 - 1)},\\
    \label{rho}  \rho &=& \textcolor{red}{-}\frac{C^{(m)}_2}{\sqrt{{C^{(m)}_3}^2 - 1}} \frac{1}{r^2},
\end{eqnarray}
where the sign $u < 0$ ($u > 0$) indicates ingoing (outgoing) fluid. In the present study, we focus on the accretion process ($u < 0$). However, for visual convenience, Fig. \ref{radial} displays the magnitude of the radial velocity $|u|$ as a function of the coordinate $r$. The constants $C^{(m)}_3$ and $C^{(m)}_2$ are determined by initial conditions at the boundary $r_i = 500b_0$. For our analysis, we set $u_i = -0.5$ and $\rho_i = 0.001$.

By analyzing the plot of $|u|$, we observe that the velocity magnitude decreases as the particle approaches the throat, in agreement with the results reported in \cite{Rueda:2023val}. This behavior contrasts with the typical case of black holes \cite{Bahamonde:2015uwa}, where the radial velocity increases as $r$ decreases. However, an important feature in our case is that, as the parameter 
$m$ increases, the radial velocity decreases more gradually with decreasing 
$r$, until it eventually drops abruptly to zero near the throat for sufficiently large values of $m$.

By analyzing the expression \eqref{rho} for the energy density, we observe that its behavior does not vary significantly with the parameter 
$m$. The density falls off as 
$1/r^2$
  and reaches its maximum at the throat, suggesting that the highest concentration of matter is localized around this region, as expected. Another important point to highlight is that the energy density remains regular throughout the entire spacetime, in contrast to the typical behavior in black holes \cite{Rueda:2023val}, where singularities usually arise.

To conclude our analysis, we examine the mass transport rate $\dot{M}$ of the dust traversing the GEB wormhole, which corresponds to the flux integral given by \cite{Debnath:2015hea}
\begin{eqnarray}
    \dot{M} = \int T_0^1 \sqrt{-g}d\theta d\varphi,
\end{eqnarray}
which in our case leads to
\begin{eqnarray}
    \dot{M} = \textcolor{red}{-}4\pi C^{(m)}_2\left(1 - \frac{b_0^m}{r^m}\right)^{2 -2/m}\sqrt{u^2\left(1 - \frac{b_0^m}{r^m}\right)^{2 -2/m} +1}.
\end{eqnarray}

\begin{figure}
    \centering
    \includegraphics[width=0.5\linewidth]{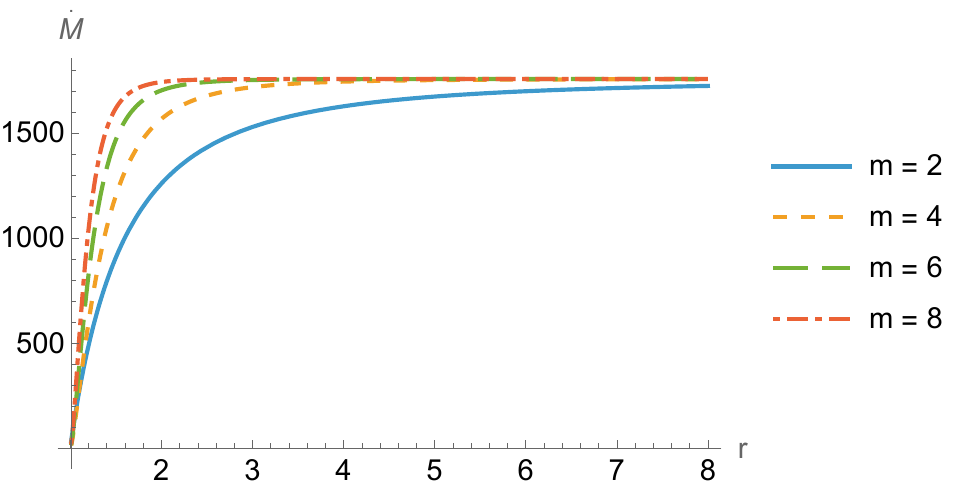}
    \caption{Magnitude of mass transport rate for different values of $m$ and fixing $b_0=1$.}
    \label{mass}
\end{figure}

 In Fig. \ref{mass} we have the plot of $|\dot{M}|$ for different values of $m$.

Observing the plot above, we see that the magnitude mass transport rate change decreases as we approach the wormhole throat, following the behavior of the radial velocity, which also decreases near the throat.

A similar analysis for the standard $m=2$ case was performed in \cite{Yusupova:2021rcz} (see also the 2026 Erratum \cite{Yusupova:2026err}). It is important to note that while their work uses a massive EB wormhole in isotropic coordinates, our GEB model with $m=2$ is exactly equivalent to theirs under the coordinate transformation $r = \tilde{r}(1 + b_0^2/4\tilde{r}^2)$, for the massless case. By accounting for this transformation and the corresponding Jacobian, our analytical results for dust match their corrected expressions. As observed in both works, the corrected mass transport rate consistently predicts a decrease in the wormhole's mass parameter during accretion.

We emphasize that a detailed study of geodesics for wormholes with an accretion disk has already been carried out in Ref.~\cite{Rueda:2023val}. In that work, a comprehensive analysis of photon trajectories under steady accretion shows that the orbits exhibit only a single maximum at the throat, with no stable photon spheres. These results are crucial in constraining shadow formation in such geometries and directly apply to the scenario discussed in this context. In the framework of GR, in the next section, we calculate the fields that serve as sources for this geometry.

\section{Field Sources for GEB wormhole}
In this section, we calculate the fields that serve as sources for this geometry. It is important to note that the source of a geometry is a matter of interpretation. For $m=2$, it has been shown that the solution can be supported by a Maxwell field and a phantom fluid \cite{Shatskiy:2008zz}. However, for the generalized $m > 2$ case, instead of resorting to the ``engineering" of a phenomenological anisotropic fluid, which would be a trivial exercise in fitting pressures, we seek a more fundamental origin. We will show that this geometry arises as an exact solution of a field theory involving NED and a phantom scalar field. This approach aligns our work with modern methodologies that map geometric requirements into fundamental Lagrangians, providing a more rigorous physical basis for the GEB wormhole.

\subsection{General Relations}
Let's consider an action of the form

\begin{equation}\label{acao}
    S = \int \sqrt{-g}d^4x\left(R - 2\epsilon g^{\mu\nu}\partial_\mu\phi\partial_\nu\phi + 2V(\phi) + \mathcal{L(F)}\right),
\end{equation}
where $\phi$ is a scalar field, $V(\phi)$ is the potential associated to the
 scalar field and $\mathcal{L(F)}$ is the Lagrangian density of the nonlinear electromagnetic field with $\mathcal{F} = F^{\mu\nu}F_{\mu\nu}$, where $F_{\mu\nu}$ is the electromagnetic field tensor. Furthermore, $\epsilon = -1$ for a phantom scalar field and $\epsilon = +1$ for a canonical scalar field.

Varying the action with respect to the metric 
$g^{\mu\nu}$ yields Einstein's equation
\begin{equation}
    G_{\mu\nu}= R_{\mu\nu} - \frac{1}{2}R g_{\mu\nu} = T_{\mu\nu}[\phi] + T_{\mu\nu}[\mathcal{F}],
\end{equation}
where $T_{\mu\nu}[\phi]$ e $ T_{\mu\nu}[\mathcal{F}]$ are, respectively, the energy-momentum tensor associated with the scalar field and the nonlinear electromagnetic field, given by
\begin{equation}
    T_{\mu\nu}[\phi] = 2\epsilon\partial_\mu\phi\partial_\nu\phi - g_{\mu\nu}\left(\epsilon g^{\lambda\rho}\partial_\lambda\phi\partial_\rho\phi - V(\phi)\right),
\end{equation}
\begin{equation}
    T_{\mu\nu}[\mathcal{F}] = g_{\mu\nu}\frac{\mathcal{L(F)}}{2} - 2\mathcal{L_F}F^{\alpha}_\mu F_{\nu\alpha},
\end{equation}
where$\mathcal{L_F} = d\mathcal{L}/d\mathcal{F}$. 
Varying the action with respect to the scalar field $\phi$ and the electromagnetic field $F_{\mu\nu}$ we obtain the equations for the fields
\begin{equation}\label{fieldeq}
    2\epsilon\nabla_\mu\nabla^\mu\phi = - \frac{dV(\phi)}{d\phi},
\end{equation}
\begin{equation}\label{Feq}
    \nabla_\mu (\mathcal{L_F} F^{\mu\nu}) = 0.
\end{equation}

\subsection{Magnetic Source}

In addition to assuming a radial scalar field $\phi = \phi(r)$, we also assume here only the existence of a radial magnetic field, i.e., that the only non-zero components of the electromagnetic tensor $F_{\mu\nu}$ are given by $F_{\theta\varphi} = - F_{\varphi\theta} = q_m\sin\theta$, where $q_m$ is the charge of the magnetic monopole. With this, we have that the invariant $\mathcal{F}$ is given by
\begin{equation}\label{F}
    \mathcal{F} = \frac{2q_m^2}{r^4}.
\end{equation}

With the metric given in \ref{metrica}, the components of the energy momentum  tensors for $\phi$ and $F_{\mu\nu}$ take the form
\begin{equation}
    T_\mu^\nu [\phi] = V(\phi) \delta_\mu^\nu - \epsilon\left(1 - \frac{b(r)}{r}\right)\phi'(r)^2\text{diag(1,-1 ,1 ,1)},
\end{equation}
\begin{equation}
    T_\mu^\nu[\mathcal{F}] = \frac{1}{2}\text{diag}\left(\mathcal{L}, \mathcal{L}, \mathcal{L} - \frac{4q_m^2}{r^4}\mathcal{L_F}, \mathcal{L} - \frac{4q_m^2}{r^4}\mathcal{L_F}\right).
\end{equation}

Looking at the above equations, we see that $G^t_t - G^r_r = T^t_t - T^r_r$ is free of $V$ and $\mathcal{L}$, we have that, using the components of Einstein's equation given in \eqref{Et} e \eqref{Er}
\begin{equation}
    -2\epsilon\left(1 - \frac{b(r)}{r}\right)\phi'(r)^2 = \frac{b(r)}{r^3} - \frac{b'(r)}{r^2}.
\end{equation}

Using the equation \eqref{b}, we finally have that the field is given by
\begin{equation}\label{phieq}
    \phi'(r)^2 = \frac{(m - 1)}{\epsilon r^2\left(1 - r^m/b_0^m\right)}.
\end{equation}

As $r > b_0$, for $\phi'(r)^2 > 0$, we must have $\epsilon = -1$, that is, a phantom scalar field. Solving the equation \eqref{phieq}, we find
\begin{equation}\label{phi}
    \phi(r) = \phi_0 + 2\frac{\sqrt{m - 1}}{m}\arctan\left(\frac{\sqrt{r^m - b_0^m}}{b_0^{m/2}}\right),
\end{equation}
or, in terms of $l$
\begin{equation}
    \phi(l) = \phi_0 + 2\frac{\sqrt{m - 1}}{m}\arctan\left(\frac{l}{b_0}\right)^{m/2},
\end{equation}
where $\phi_0$ is a constant that we can fix as zero. From the equation \eqref{phi}, it is simple to see that for $m = 2$ we have the scalar field for the Ellis-Bronnikov case \cite{Bronnikov:2018vbs}. The scalar field \eqref{phi} is quite similar to the scalar field that serves as the Simpson-Visser spacetime source \cite{Bronnikov:2021uta}, which is also an arc-tangent function, a functional form that is quite recurrent in singularity-free space-times.

%\begin{figure}[h]
    %\centering
    %\includegraphics[width=0.7\linewidth]{phi.pdf}
   % \caption{Plot do campo escalar $\phi(r)$ para diferentes valores de $m$, com $b_0 = 1$ e $\phi_0 = 0$.}
  %  \label{fig:enter-label}
%\end{figure}

Using the equation \eqref{fieldeq}, we can integrate to find the potential $V(r)$, given by
\begin{equation}
    V(r) = \frac{(m - 1)(m - 2)b_0^m}{m r^{2m}}(r^m - b_0^m)^{1 - 2/m} + \frac{2(m - 1)(r^m - b_0^m)^{1 - 2/m}}{m b_0^m}\, _2F_1\left[2, 1 - \frac{2}{m}, 2 - \frac{2}{m}, 1 - \frac{r^m}{b_0^m}\right],
\end{equation}
where $_2F_1(a,b,c, z)$ is the hypergeometric function of $z$. Using the equation \eqref{phi} to invert, we have that the potential written in terms of the $\phi$ field is given by

\begin{align}\label{v}
    V(\phi) =& \frac{(m -1)}{mb_0^2}\tan^{2 - 4/m}\left(\frac{m\phi}{2\sqrt{m - 1}}\right)\left\{(m -2)\cos^4\left(\frac{m\phi}{2\sqrt{m - 1}}\right) \right.\nonumber \\
    &\left.+ 2 \,\,_2F_1\left[2, 1 - \frac{2}{m}, 2 - \frac{2}{m}, -\tan^2\left(\frac{m\phi}{2\sqrt{m - 1}}\right)\right]\right\}
\end{align}

From Einstein's equation $(t,t)$ $G^t_t = T^t_t$, we have
\begin{equation}
    -\frac{b'(r)}{r^2}= V(r) + \left(1 - \frac{b(r)}{r}\right)\phi'(r)^2 + \frac{\mathcal{L}(r)}{2},
\end{equation}
or
\begin{equation}
    \frac{\mathcal{L}(r)}{2} = - \frac{b'(r)}{r^2} - V(r) -  \left(1 - \frac{b(r)}{r}\right)\phi'(r)^2.
\end{equation}

Using the expressions for $\phi(r)$ and $V(r)$ found, we obtain the Lagrangian in terms of $r$
\begin{align} 
    \mathcal{L}(r) =& \frac{2}{(r^m - b_0^m)^{2/m}}\left\{1 - \frac{(m - 2)}{m}\left(\frac{b_0}{r}\right)^{2m} - \frac{2}{m}\left(\frac{b_0}{r}\right)^{m} - \frac{(r^m - b_0^m)^{2/m}}{r^2} \right.\nonumber \\
    &\left.- \frac{2(m - 1)(r^m - b_0^m)}{m b_0^m} \,\, _2F_1\left(2, 1 - \frac{2}{m}, 2 - \frac{2}{m}, 1 - \frac{r^m}{b_0^m}\right)\right\}.
\end{align}

\begin{figure}[h!]
    \centering
    \includegraphics[width=0.5\linewidth]{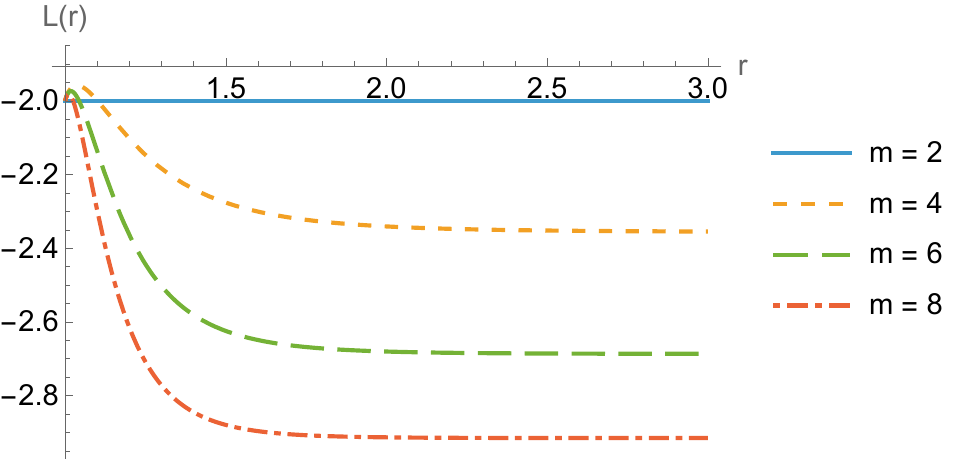}
    \caption{Lagrangian $L(r)$ for magnetic case for different values of $m$, with $b_0 = 1$.}
    \label{L(r)Mag}
\end{figure}

In Fig.\ref{L(r)Mag} , we have the plot of the Lagrangian as a function of $r$, where we can see a similar behavior for all $m > 2$, in which $\mathcal{L}(r) \to $ constant for $r \to \infty$.

We can use the equation \eqref{F} to invert the Lagrangian equation and write it in terms of the invariant $\mathcal{F}$, from which we get
\begin{align}
        \mathcal{L(F)} =& \frac{2}{b_0^2[(s \mathcal{F}^{-1/4})^m - 1]^{2/m}}\left\{1 - \frac{(m - 2)}{m}\left(\frac{\mathcal{F}^{1/2}}{s^2}\right)^m - \frac{2}{m}\left(\frac{\mathcal{F}^{1/4}}{s}\right)^m - \frac{\mathcal{F}^{1/2}}{s^2}\left[\left(\frac{s}{\mathcal{F}^{1/4}}\right)^m - 1\right]^{2/m}\right.\nonumber\\
        & \left. + \frac{2(m - 1)}{m}\left[1 - \left(\frac{s}{\mathcal{F}^{1/4}}\right)^m\right]\,\, _2F_1\left(2, 1 - \frac{2}{m}, 2 - \frac{2}{m}, 1 - \left(\frac{s}{\mathcal{F}^{1/4}}\right)^m\right)\right\},
\end{align}
where
\begin{equation}
    s = \frac{(2q_m^2)^{1/4}}{b_0}.
\end{equation}

As a simplification, we can also define the quantity
\begin{equation}
    \mathcal{G} = \frac{\mathcal{F}^{1/4}}{s},
\end{equation}
so that in terms of $\mathcal{G},$ we have
\begin{align}
    \mathcal{L(G)} =& \frac{2}{b_0^2(\mathcal{G}^{-m} - 1)^{2/m}}\left\{1 - \frac{(m - 2)}{m}\mathcal{G}^{2m} - \frac{2}{m}\mathcal{G}^m - \mathcal{G}^2(\mathcal{G}^{-m} - 1)^{2/m} \right.\nonumber\\
    &\left.+ \frac{2(m - 1)}{m}(1 - \mathcal{G}^{-m})\,\, _2F_1\left(2, 1 - \frac{2}{m}, 2 - \frac{2}{m}, 1 - \mathcal{G}^{-m}
\right)  \right\}.
\end{align}

%\begin{figure}[h]
   % \includegraphics[width=0.75\linewidth]{L(F).pdf}
  %  \caption{Caption}
  %  \label{fig:enter-label}
%\end{figure}

It is easy to see that for $m = 2$ we recovered the usual Ellis-Bronnikov case, for which the field source is just a phantom free scalar field. That is, for $m = 2$, we have $V(\phi) = 1/b_0^2$ and $\mathcal{L(F)} = -2/b_0^2$. By substituting this in the action \eqref{acao} we get
\begin{equation}
    S( m = 2) = \int d^4x \left(R - 2\epsilon g^{\mu\nu}\partial_\mu\phi\partial_\nu\phi + \frac{2}{b_0^2} - \frac{2}{b_0^2}\right) = \int d^4x \left(R - 2\epsilon g^{\mu\nu}\partial_\mu\phi\partial_\nu\phi \right),
\end{equation}
that is, for $m = 2$, the action reduces to the action of a free scalar field, as expected.

\subsection{Electric Source}

Alternatively, instead of considering a radial magnetic field generated by a magnetic monopole, we can consider a radial electric field, so that now the non-zero components of the electromagnetic tensor $F_{\mu\nu}$ are given by $F^{tr} = - F^{rt} = E(r)$. With this, the invariant $\mathcal{F}$ is now given by 
\begin{equation}\label{E}
    \mathcal{F} = -\frac{2E(r)^2}{\left(1 - \frac{b(r)}{r}\right)}.
\end{equation}

Also, using the equation \eqref{Feq} we can write the electric field as
\begin{equation}\label{E(r)}
    E(r) = \frac{q_e}{r^2\mathcal{L_F}}\sqrt{1 - \frac{b(r)}{r}},
\end{equation}
where $q_e$ is the electric charge.

With this, the tensor energy momentum associated with NED is given by
\begin{equation}
    T[\mathcal{F}]^\nu_\mu = \frac{1}{2}\text{diag}\left(\mathcal{L} + \frac{4E(r)^2\mathcal{L_F}}{1 - \frac{b(r)}{r}}, \mathcal{L} + \frac{4E(r)^2\mathcal{L_F}}{1 - \frac{b(r)}{r}}\mathcal{L}, \mathcal{L}\right)
\end{equation}

Here we proceed analogously to the magnetic case to find the scalar field and the associated potential, which are given by the same expressions \eqref{phi} and \eqref{v} as the magnetic case.

Finally, from Einstein's equation $(\theta, \theta)$ we get
\begin{equation}
    L(r) = \frac{b(r) - rb'(r)}{2r^3} - V(r) - \left(1 - \frac{b(r)}{r}\right)\phi'(r)^2.
\end{equation}

Finally, substituting $\phi(r)$ and $V(r)$, we find that the Lagrangian is given by
\begin{equation}
    \mathcal{L}(r) = -2\frac{(m - 1)(m - 2)b_0^m(r^m - b_0^m)^{1 - 2/m} }{m r^{2m}}- \frac{4(m - 1)(r^m - b_0^m)^{1 - 2/m}}{m b_0^m}\, _2F_1\left[2, 1 - \frac{2}{m}, 2 - \frac{2}{m}, 1 - \frac{r^m}{b_0^m}\right].
\end{equation}

Finally, from the equation $(r,r)$ we can solve for $\mathcal{L}_F$, since the electric field in terms of $\mathcal{L}_F$ is given in \eqref{E(r)}. By doing this, we get
\begin{equation}
    \mathcal{L_F}(r) = -\frac{2q_e^2}{r^2[1 - r^{2 - 2m}(r^m + (m - 2)b_0^m)(r^m - b_0^m)^{1 - 2/m}]}.
\end{equation}

\begin{figure}[h]
    \centering
    \includegraphics[width=0.5\linewidth]{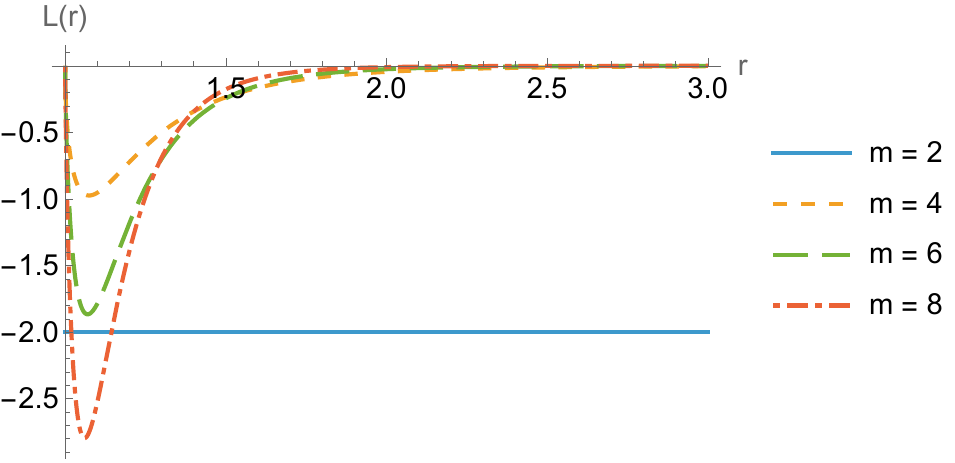}
    \caption{Lagrangian $\mathcal{L}(r)$ for electric case for different values of $m$, with $b_0 = 1$.}
    \label{L(r)ele}
\end{figure}

In Fig.\ref{L(r)ele} we have the plot of the Lagrangian $\mathcal{L}$ as a function of $r$. It is worth noting that, as usually happens in the electrical case, it is not possible to write $\mathcal{L}$ as an explicit function of $\mathcal{F}$, since $\mathcal{F}$ depends on a non-trivial form of $r$. In addition, it is possible to see that, as in the magnetic case, the Lagrangian tends to a constant for asymptotic values of $r$; For small values of $r$, $\mathcal{L}(r)$ tends to a constant to $r \to b_0$.

Finally, the expression for the electric field is given by
\begin{equation}
    E(r) = \frac{(r^m -b_0^m)^{1 - 3/m}}{2q_er^{m + 1}}\left[r^4 - b_0^{2m}(m - 2)r^{4 - 2m} + b_0^m(m - 3)r^{4 - m} - r^2(r^m - b_0^m)^{2/m}\right].
\end{equation}

\begin{figure}[h]
    \centering
    \includegraphics[width=0.5\linewidth]{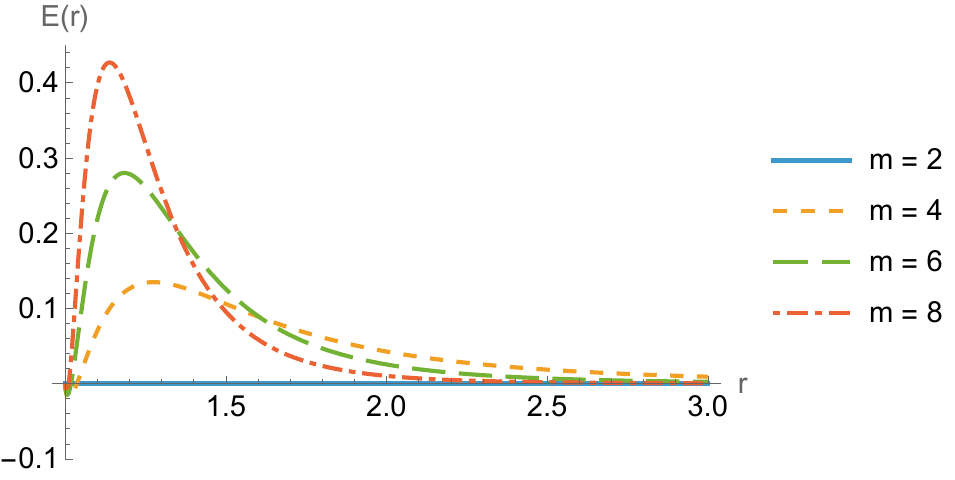}
    \caption{Electric field for different values of $m$, with $b_0 = 1$.}
    \label{E(r)fig}
\end{figure}

This electric field is obviously not Coulombian. On the limit $r \to \infty$
\begin{equation}
    E(r) \sim \left(\frac{b_0}{r}\right)^m,
\end{equation}
which demonstrates that the decay of the $E(r)$ field is faster the higher the value of $m$, which is in agreement with the diagram in Fig.\ref{embebed}, which shows that the geometry tends to be flat faster as we increase the value of $m$.
In the graph of Fig.\ref{E(r)fig} we have the behavior of the electric field for different values of $m$, where we see that, for $r \to b_0$, the electric field does not have a divergent behaviour.

\section{Conclusion}

In this work, we explore the characteristics and properties of the GEB wormhole within the framework of GR. We first analyze how the embedding diagrams are affected by the parameter $m$, and then briefly study the accretion of dust in this spacetime. It is shown that, unlike what typically occurs in black holes, the dust radial velocity decreases as the particle approaches the throat, and this decrease becomes increasingly abrupt as the parameter 
$m$ increases. Regarding the energy density, there is a higher concentration near the wormhole throat. Furthermore, we show that the mass transport rate it is always negative, indicating a loss of mass from the wormhole.

In this context, it is important to discuss the physical viability of this accretion analysis given the known instabilities of scalar-supported wormhole geometries. It is well documented that the standard Ellis-Bronnikov family exhibits instabilities under both linear and non-linear perturbations \cite{Gonzalez:2008wd, Gonzalez:2008xk, Nandi:2016ccg}. Therefore, our analysis is developed under the \textit{assumption} of a quasi-stable regime, acting as a working hypothesis where the dynamical timescale of the infalling dust $\tau_{\text{acc}}$ is assumed to be sufficiently separated from the instability growth timescale $\tau_{\text{inst}}$. We acknowledge, however, that in standard EB geometries, these timescales are often of the same order, and a clear separation is not generically guaranteed. Furthermore, while the presence of the NED sector required for the $m > 2$ generalization might potentially alter the stability landscape, such a possibility remains speculative at this stage, as a formal stability analysis for these specific configurations has not yet been performed. The characteristic accumulation of matter near the throat, driven by the abrupt deceleration of the flow, raises compelling questions regarding back-reaction effects. Whether this accumulated mass accelerates gravitational collapse or contributes to a dynamical stabilization is a non-trivial open problem that warrants future rigorous investigation.

We have also explored whether scalar fields coupled with NED can serve as field sources for GEB spacetimes. By employing both magnetic and electric sources in the context of NED theories, we derived analytical expressions for the scalar field, its corresponding potential, and the NED Lagrangian. Notably, for the magnetic solution, we successfully inverted the equations, expressing the Lagrangian as a function of the electromagnetic invariant \( \mathcal{F} = F_{\mu\nu}F^{\mu\nu} \). A key observation is that for \( m = 2 \), the solution recovers the well-known EB spacetime sourced by a free phantom scalar field. This connection highlights the versatility of the arc-tangent scalar field, which frequently arises in singularity-free spacetimes, including those of the Simpson-Visser class.

For the electric case, the decay of the electric field is influenced by the parameter \( m \), with higher values of \( m \) leading to a faster approach to flat spacetime, as demonstrated in our results. These distinctions between the magnetic and electric cases offer insight into the behavior of fields in NED-modified spacetimes, emphasizing that the source structure plays a critical role in shaping spacetime geometry.

The ability to express the Lagrangian in terms of the electromagnetic invariant \( \mathcal{F} \) for the magnetic case but not for the electric one, due to the dependence of \( \mathcal{F} \) on \( r \), highlights the richness and complexity of NED as a field source for wormholes. These results not only advance our understanding of the scalar-NED interplay but also open new avenues for further research into different classes of wormhole geometries. Specifically, the approach presented here introduces a novel methodology for seeking field sources in more generalized or modified wormhole spacetimes using NED and scalar fields as core ingredients.

Future investigations may explore whether the class of solutions presented here remains stable under perturbations. In particular, linear stability analyses — such as quasinormal mode studies — and nonlinear time-evolution approaches would provide valuable insights into the robustness of the configuration. Additionally, observational aspects, such as gravitational wave signatures or matter dynamics near the throat, are promising directions. The framework developed here serves as a path for exploring a broader range of NED-modified geometries and for the continued search for non-singular, physically viable solutions to wormhole spacetimes.

\section*{Acknowledgements}

Tiago M. Crispim, G. Alencar and Celio R. Muniz would like to thank Conselho Nacional de Desenvolvimento Científico e Tecnológico (CNPq) and Fundação Cearense de Apoio ao Desenvolvimento Científico e Tecnológico (FUNCAP) for the financial support and Marcos Silva for  valuable discussions and insightful comments on the manuscript

\end{document}